\documentclass[aps,prl,twocolumn,superscriptaddress,groupedaddress]{revtex4}

\usepackage{amssymb}
\usepackage{color,graphicx}
\usepackage{amsmath}
\usepackage{amsbsy}
\usepackage{amsthm}
\usepackage{bbm}
\usepackage{bm}
\usepackage{epsfig}
\usepackage{lscape}
\usepackage{float}
\usepackage{subfigure}

\newcommand{\D}{\text{d}}

\begin{document}

\title{Is Gravity Quantum?}

\author{M. Bahrami}
%\email{Mohammad.Bahrami@ts.infn.it}
\affiliation{Department of Physics, University of Trieste, Strada Costiera 11, 34014 Trieste, Italy}
\affiliation{Istituto
Nazionale di Fisica Nucleare, Trieste Section, Via Valerio 2, 34127 Trieste,
Italy}

\author{A. Bassi}
%\email{bassi@ts.infn.it}
\affiliation{Department of Physics, University of Trieste, Strada Costiera 11, 34014 Trieste, Italy}
\affiliation{Istituto
Nazionale di Fisica Nucleare, Trieste Section, Via Valerio 2, 34127 Trieste,
Italy}

\author{S. McMillen}
%\email{m.paternostro@qub.ac.uk}
\affiliation{Centre for Theoretical Atomic, Molecular and Optical Physics, School of Mathematics and Physics, Queen\textquoteright{}s University, Belfast BT7 1NN, United Kingdom}

\author{M. Paternostro}
%\email{m.paternostro@qub.ac.uk}
\affiliation{Centre for Theoretical Atomic, Molecular and Optical Physics, School of Mathematics and Physics, Queen\textquoteright{}s University, Belfast BT7 1NN, United Kingdom}

\author{H. Ulbricht}
%\email{h.ulbricht@soton.ac.uk}
\affiliation{School of Physics and Astronomy, University of Southampton, Southampton SO17 1BJ, United Kingdom}

\date{\today}

\begin{abstract}
What gravitational field is generated by a massive quantum system in a spatial superposition? This is one of the most important questions in modern physics, and after decades of intensive theoretical and experimental research, we still do not know the answer. On the experimental side, the difficulty lies in the fact that gravity is weak and requires large masses to be detectable. But for large masses, it becomes increasingly difficult to  generate spatial quantum superpositions, which live sufficiently long to be detected. A delicate balance between opposite quantum and gravitational demands is needed. Here we show that this can be achieved in an optomechanics scenario. We propose an experimental setup, which allows to decide whether the gravitational field generated by a quantum system in a spatial superposition is the superposition of the two alternatives, or not. We estimate the magnitude of the effect and show that it offers good perspectives for observability. Performing the experiment will mark a breakthrough in our understanding of the relationship between gravity and quantum theory.
\end{abstract}

\pacs{xxxxxxxxxxxx}

\maketitle

%%%%%%%%%%%%%%%%%%%%%%%%%%%%%%%%%%%%%%%%%%%%%%%%%%%%%%%%%%%%%%%%%%%%%%%%%%%%%%%%%%%%%
Quantum field theory is one of the most successful theories ever formulated. All matter fields, together with the electromagnetic and nuclear forces, have been successfully  embodied in the quantum framework. They form the much celebrated standard model of elementary particles, which not only has been confirmed in all advanced accelerator facilities, but has also become an essential ingredient for the description of the universe and its evolution. 

In light of this, it becomes obvious to seek a quantum formulation of gravity as well. Yet, the straightforward procedure for promoting the classical field as described by general relativity, into a quantum field, does not work. Over the decades, several strategies have been put forward, which turned into very sophisticated theories of gravity, perhaps the most advanced being string theory and loop quantum gravity. Yet,  none of them has reached the goal of providing a fully consistent quantum theory of gravity.  

At this point, one might wonder whether the very idea of quantizing gravity is ill-posed~\cite{Penrose2014, Dys}. At the end of the day, according to general relativity, gravity is rather different from all other forces. Actually, it is not a force at all, but a mere manifestation of the curvature of spacetime, and there is no obvious reason why the standard approach to the quantisation of fields should work for spacetime as well.  A future unified theory of quantum and gravitational phenomena might require a radical revision not only of our notions of space and time, but also of (quantum) matter. This scenario is growing in likeliness. 

From the experimental point of view, it has now been ascertained that quantum matter (i.e. matter in a genuine quantum state, such as a coherent superposition state) couples to the Earth's gravity in the most obvious way. This has been confirmed in neutron~\cite{COW}, atom~\cite{Peters} interferometers and used for velocity selection in molecular interferometry~\cite{Brez}. However, in all cases, the gravitational field is classical, i.e. it is generated by a distribution of matter (the Earth) in a fully classical state.  Therefore, the plethora of successful experiments mentioned above does not provide hints, unfortunately, on whether gravity is quantum or not. 

The large attention and media coverage about the BICEP2 experiment having shown the quantum origin of primordial gravitational fluctuations~\cite{BICEP2}, subsequently disproved by Planck's data analysis~\cite{Planck}, testifies the importance and urgency of a pragmatic assessment of the question of whether gravity is quantum or not.

In this paper, we propose an approach where a mesoscopic system is forced in the superposition of two different positions in space, and its gravitational field is explored by a probe (Fig. 1). Using the exquisite potential for transduction officered by optomechanics, we can in principle determine whether the gravitational field is the superposition of the two gravitational fields associated to the two different states of the system, or not. The first case amounts to a quantum behavior of gravity, the second to a classical-like one. We show that the sensitivity necessary to appreciate the difference between such behaviors is close to the current state of the art in specific optomechanical configuration, although quite demanding.

\begin{figure}[t!]
\centering
\includegraphics[width=0.6\columnwidth]{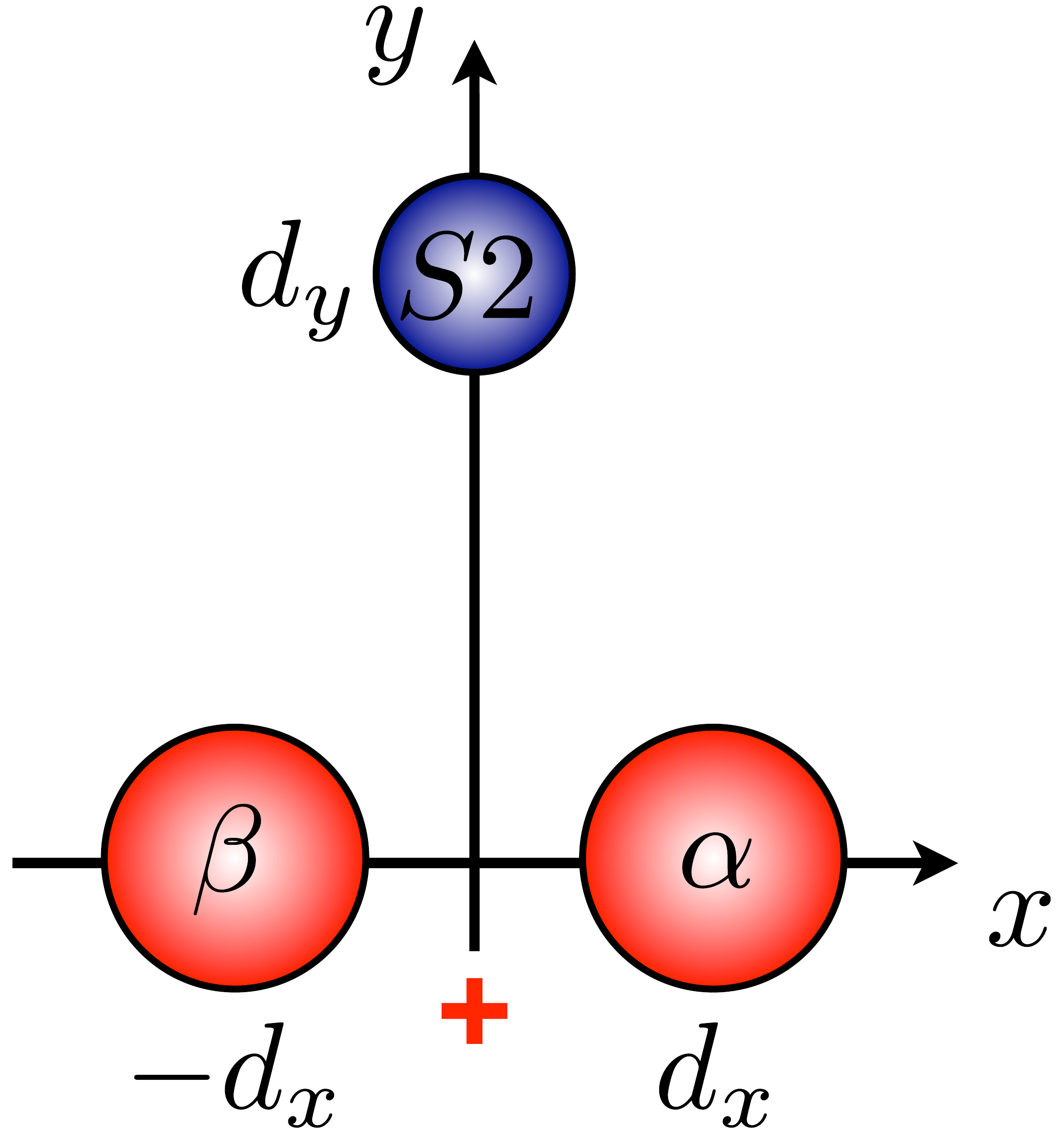}
\caption{{A two-body system that interacts only gravitationally}. To study how these systems interact gravitationally, we only need to consider the degrees of freedom in a two-dimensional $xy$-plane. System S1 is initially prepared in a superposition of two wavepackets delocalised along the $x$ axis of the reference frame (red balls). System S2 (the probing system) is initially prepared in a narrowly localized wavepacket (blue ball). The parameters of the experiment are adjusted in such a way that the non-gravitational interactions between S1 and S2 are, for all practical purposes, negligible.
}
\label{fig1}
\end{figure}

\noindent
{\it Framework.--} We consider a system S1 (with mass $m_1$) prepared in a superposition of two different positions in space. The wave function is $\psi({\bf r}_1)=\frac{1}{\sqrt{2}}\left(\alpha({\bf r}_1)+\beta({\bf r}_1)\right)$ with $\gamma({\bf r}_1)=\langle{\bf r}_1|\gamma\rangle~(\gamma=\alpha,\beta)$ and $\langle\alpha|\beta\rangle=0$, stating the distinguishability (in a macroscopic sense) of the two  states. S1  generates a gravitational field that can be probed with the help of a second system S2 (with mass $m_2$). The latter is prepared in a localized state with associated wave function $\phi({\bf r}_2)$ [cf. Fig.~\eqref{fig1}]. The question  we address is: which gravitational field does S2 experience, as generated by S1 being in a spatial superposition? We consider two alternatives. 

% For a quantum superposition of gravitational fields, S2 will move either to the right or to the left, along the axis of the spatial superposition. However, in a classical formulation of gravity, the gravitational field of S1 is most-likely connected to its wavefunction, which would embody a sort of mass density. Thus, S2 would feel a gravitational field produced by a system with mass density $m_1|\psi({\bf r}_1)|^2$, thus giving rise to an evolution of S2 along the axis of the spatial superposition that will be profoundly different from the previous case. %Applying this scenario in an optomechanical experimental setting, we propose an experiment that can be used to testing whether gravity is quantum or classical at the fundamental level.

%%%%%%%%%%%%%%%%%%%%%%%%%%%%%%%%%%%%%%%%%%%%%%%%%%%%%%%%%%%%%%%%%%%%%%%%%%%%%%%%%%%%%5
\noindent {\it Quantum gravity scenario}.--
%%%%%%%%%%%%%%%%%%%%%%%%%%%%%%%%%%%%%%%%%%%%%%%%%%%%%%%%%%%%%%%%%%%%%%%%%%%%%%%%%%%%%5
Although we do not yet have a fully consistent and complete theory of quantum gravity, we can safely claim that, should gravity be quantum,  this would be manifested in S1 generating the superposition of two weak gravitational fields, one associated to the state $\alpha({\bf r}_1)$ and the other one to the state $\beta({\bf r}_1)$. Linearity is the characteristic trait of quantum theory, which should find place in any quantization recipe for gravity. On top of this, nonlinear quantum theories very easily run into serious troubles~\cite{Wei, Pol, Gis}. Therefore S1 generating the superposition of two gravitational fields is the most likely scenario in a world where also gravity is quantum.

Then S2 reacts to the situation described here above by turning into a superposition of being attracted towards  the region $A$ where $\alpha({\bf r}_1)$  is different from zero, and being attracted towards $B$, where $\beta({\bf r}_1)$ is different from zero. If the motion of S2 is constrained within an harmonic trap, a position measurement will reveal it being  displaced slightly away from equilibrium, alternatively closer to $A$ or $B$.

% S2 feeling a gravitational field resulting from the linear combination of the contributions arising from the superposition state of S1 (see Fig.\eqref{fig1}). This linearity feature is the characteristic trait of quantum theory, which has remained valid when standard quantum theory has been extended to other realms, such as quantum electrodynamics or the standard model. In this regard, by \emph{quantum gravity}, we intend here a theory in which, at least at the weak-field, non-relativistic regime, the gravitational interaction reads as the Newtonian gravitational interaction~\cite{Bahrami_SN}, in the same sense that quantum electrodynamics implies the Coulomb interaction at the low-energy limit.

%%%%%%%%%%%%%%%%%%%%%%%%%%%%%%%%%%%%%%%%%%%%%%%%%%%%%%%%%%%%%%%%%%%%%%%%%%%%%%%%%%%%%5
\noindent {\it Semiclassical gravity scenario}.--
%%%%%%%%%%%%%%%%%%%%%%%%%%%%%%%%%%%%%%%%%%%%%%%%%%%%%%%%%%%%%%%%%%%%%%%%%%%%%%%%%%%%%5
Suppose instead that gravity, for whatever reason, is fundamentally classical. In this case, no one really knows which gravitational field S1 generates. However, a natural answer is that the (square modulus of the) wave function acts as some sort of matter distribution, which generates the gravitational field. This is what  semi-classical Einstein equations~\cite{Moller_Rosenfeld,isham,Bahrami_SN} predict, if taken seriously as fundamental equations. It is also compatible with our current knowledge of gravity. 

In this case, the gravitational field is not the quantum superposition of the two quantum fields associated to $\alpha({\bf r}_1)$ and $\beta({\bf r}_1)$, but the classical sum of the two classical fields generated by their square modulus, respectively. In such conditions, S2 feels a force which pulls it in between region $A$ and $B$. Quite evidently, the two cases imply two different motions for the probe S2, such difference being the way to discriminate between classical-like and quantum treatment. 

% Treating gravity classically, the source of the gravitational field is reasonably given by the squared modulus of the wavefunction. This is, in fact, the case in semi-classical Einstein equations~\cite{Bahrami_SN}. Accordingly, in the classical approach, the gravitational field of S1 is the gravitational field produced by a system whose mass density is $m_1|\psi({\bf r}_1)|^2$.

% In brief, quantum gravity predicts that the gravitational field of S1 is a linear combination of the gravitational fields produced by different terms of the superposition, while such superposed gravitational fields do not exist in a classical treatment of gravity. We shall use this substantial difference to propose an experiment that is able to distinguish between classical and quantum treatments of gravity and falsifying one of the two.

%One easily realizes that the two cases imply two different motions for the probe S2. The difference is very small, but the high sensitivity of optomechanics allows to amplify and measure it {\bf Adjust sentence to what really can be done}. We now provide a quantitative estimate. 

%%%%%%%%%%%%%%%%%%%%%%%%%%%%%%%%%%%%%%%%%%%%%%%%%%%%%%%%%%%%%%%%%%%%%%%%%%%%%%%%%%%%%5
\noindent {\it Theoretical modelling}.--
%%%%%%%%%%%%%%%%%%%%%%%%%%%%%%%%%%%%%%%%%%%%%%%%%%%%%%%%%%%%%%%%%%%%%%%%%%%%%%%%%%%%%5
We refer again to the situation illustrated in Fig.\ref{fig1}. In what follows, when no explicit time dependence is reported, we imply $t=0$. We  let the two systems interact gravitationally for a time $\tau$, and then measure the position of S2 along the $x$-axis. 
The time-scale $\tau$ can be, at most, the lifetime of the superposition state $\psi$. Also, experimental parameters are adjusted such that all interactions, except gravity, are negligible, for all practical purposes.

In the quantum approach, the total Hamiltonian is given by ${H}={H}_1+{H}_2+ V_{12}({\bf r}_1-{\bf r}_2)$, where
$ H_{1,2}$ are the Hamiltonians of S1 and S2, respectively, and
$ V_{12}({\bf r}_1\!-\!{\bf r}_2)$ is the Newtonian interaction. 
The final state of the overall system is given by $\Psi({\bf r}_1,{\bf r}_2,\tau)\!=\!\left[
\Psi_\alpha({\bf r}_1,{\bf r}_2,\tau)+\Psi_\beta({\bf r}_1,{\bf r}_2,\tau)\right]/\sqrt{2}$, where each term is the solution of the equation $i\hbar\,\partial_t\,\Psi_{\alpha,\beta}\!=\!{H}\,\Psi_{\alpha,\beta}$ with initial conditions $\Psi_{\alpha}({\bf r}_1,{\bf r}_2)\!=\!\alpha({\bf r}_1)\phi({\bf r}_2)$ and $\Psi_{\beta}({\bf r}_1,{\bf r}_2)\!=\!\beta({\bf r}_1)\phi({\bf r}_2)$. We assume $m_1\gg m_2$, implying an adiabatic approximation in which the degrees of freedom of the two systems can be separated as 
\begin{equation}
%\begin{aligned}
\Psi_{\gamma}({\bf r}_1,{\bf r}_2,t)=\gamma({\bf r}_1,t)\phi_\gamma({\bf r}_2,t),~~~~~~~~(\gamma=\alpha,\beta),
%\Psi_{\beta}({\bf r}_1,{\bf r}_2,t)&=\beta({\bf r}_1,t)\,
%\phi_\beta({\bf r}_2,t),
%\end{aligned}
\end{equation}
where the motion of $\alpha({\bf r}_1,t)$ and $\beta({\bf r}_1,t)$ are determined by ${H}_1$, while $\phi_{\gamma}({\bf r}_2,t)$ evolves with the Hamiltonian ${H}_{\gamma}={H}_2+V_{\gamma}$ with
\begin{equation}
%\begin{aligned}
\label{V-qm}
 V_\gamma = -G m_1 m_2\,\int\D^3{\bf r}_1\frac{|\gamma({\bf r}_1,t)|^2}{|{\bf r}_1-{\bf r}_2|},~~~~~(\gamma=\alpha,\beta).
%V_\beta &= -G m_1 m_2\,\int\D^3{\bf r}_1\,
%\frac{|\beta({\bf r}_1,t)|^2}{|{\bf r}_1-{\bf r}_2|}.
%\end{aligned}
\end{equation}
In this quantum scenario, the initial superposition state of S1 generates a superposition of gravitational fields, which in turn generates a superposition of motions for S2.
% In fact,  each term of the superposition feels a different gravitational field. This implies the quantum character of gravity: a superposition state produces a  superposition of gravitational fields.

On the other hand, in the  semiclassical treatment of gravity, and under the same approximations discussed above, the evolution of S2 is determined by ${H}_{\text{\tiny cl}}={H}_2+ V_{\text{\tiny cl}}$,
where the gravitational potential now reads
\begin{align}\label{V-class}
V_{\text{\tiny cl}}=-Gm_1\,m_2\int\D^3{\bf r}_1\frac{|\psi({\bf r}_1,t)|^2}{|{\bf r}_1-{\bf r}_2|}.
\end{align}
Here, the evolution of $\psi({\bf r}_1,t)$ is determined by ${H}_1$.
Eq.~\eqref{V-class} shows the signatures of a classical treatment of gravity: the wave function gives the mass density, and there is no quantum superposition of gravitational fields. 

One can further approximate the above mentioned gravitational potentials, which will be quite useful when we compute the motion of optomechanical systems. 
Henceforth we will work in the Heisenberg picture. 
We assume that the quantum fluctuations around the mean values for S1 are small. %That is to say $\langle{\bf r}^2_1(t)\rangle_{\gamma}\ll \langle{\bf r}_1(t)\rangle_{\gamma}^2$, with obvious interpretation of the notation being used. % with $\langle{\bf r}_1(t)\rangle_\alpha=\langle\alpha|{\bf r}_1(t)|\alpha\rangle$. 
%We will denote $\langle{\bf r}_1(t)\rangle_\alpha$ by $\bar{\bf r}^\alpha_1(t)$.
Therefore, $ V_{\gamma}$ in Eq.~(\ref{V-qm}) can be approximated as 
\begin{equation}
%\begin{aligned}
 V_{\gamma} \approx -\frac{G m_1 m_2}{|\langle{\bf r}_1(t)\rangle_{\gamma}-{\bf r}_2(t)|},~~~~~~(\gamma=\alpha,\beta).%,\\
%V_{\beta} &\approx -\frac{G m_1 m_2}{|\langle{\bf r}_1(t)\rangle_{\beta}-{\bf r}_2(t)|}\,.
%\end{aligned}
\end{equation}
Assuming that the quantum fluctuations around the mean values for S2 are also small, %(i.e., $\langle{\bf r}^2_2(t)\rangle_\phi\ll \langle{\bf r}_2(t)\rangle_\phi^2$), then 
$ V_{\gamma}$ can be expanded in Taylor series as
\begin{equation}
\begin{aligned}
 V_{\gamma} \approx& -\frac{G m_1 m_2}{|\langle{\bf r}_1(t)\rangle_{\gamma}-
\langle{\bf r}_2(t)\rangle_\phi|}\\
&+\delta {\bf r}_2(t)\cdot\frac{G m_1 m_2(\langle{\bf r}_1(t)\rangle_{\gamma}-\langle{\bf r}_2(t)\rangle_\phi)}
{|\langle{\bf r}_1(t)\rangle_{\gamma}-\langle{\bf r}_2(t)\rangle_\phi|^3},
%\\
%V_{\beta} \approx& -\frac{G m_1 m_2}{|\langle{\bf r}_1(t)\rangle_{\beta}-
%\langle{\bf r}_2(t)\rangle_\phi|}\\
%&+\delta {\bf r}_2(t)\cdot\frac{G m_1 m_2(\langle{\bf r}_1(t)\rangle_{\beta}-\langle{\bf r}_2(t)\rangle_\phi)}
%{|\langle{\bf r}_1(t)\rangle_{\beta}-\langle{\bf r}_2(t)\rangle_\phi|^3}.
\end{aligned}
\end{equation}
where $\delta {\bf r}_2(t)={\bf r}_2(t)-\langle{\bf r}_2(t)\rangle_\phi$.

The same procedure can applied to $V_{\text{\tiny cl}}$.
As for the quantum case, assuming that the fluctuations in the motions of S1 and S2 are small, we find
\begin{align}\label{V-class-app}
\nonumber
V_{\text{\tiny cl}} &\approx 
\sum_{\gamma=\alpha,\beta}\left[-\frac{G m_1 m_2}{2|\langle{\bf r}_1(t)\rangle_\gamma-\langle{\bf r}_2(t)\rangle_\phi|}\right.\\
%-\frac{G m_1 m_2}{2|\langle{\bf r}_1(t)\rangle_\beta-\langle{\bf r}_2(t)\rangle_\phi|}
%\\ \nonumber &+
&\left.+\delta {\bf r}_2(t)\cdot\frac{G m_1 m_2(\langle{\bf r}_1(t)\rangle_\gamma-\langle{\bf r}_2(t)\rangle_\phi)}
{2|\langle{\bf r}_1(t)\rangle_\gamma-\langle{\bf r}_2(t)\rangle_\phi|^3}\right].
%\\ &+
%\delta {\bf r}_2(t)\cdot\frac{G m_1 m_2(\langle{\bf r}_1(t)\rangle_\beta-\langle{\bf r}_2(t)\rangle_\phi)}
%{2|\langle{\bf r}_1(t)\rangle_\beta-\langle{\bf r}_2(t)\rangle_\phi|^3}.
\end{align}

%In order to compute $\langle{\bf r}_2(t)\rangle_{\phi}}$, we use Ehrenfest theorem, implying $\langle{\bf r}_2(t)\rangle_{\phi}\approx(0,\bar{y}_2(t),0)$, where

\begin{figure}[t!]
\hskip0.17\columnwidth{\bf (a)}\hskip0.4\columnwidth{\bf (b)}
\includegraphics[width=0.65\columnwidth]{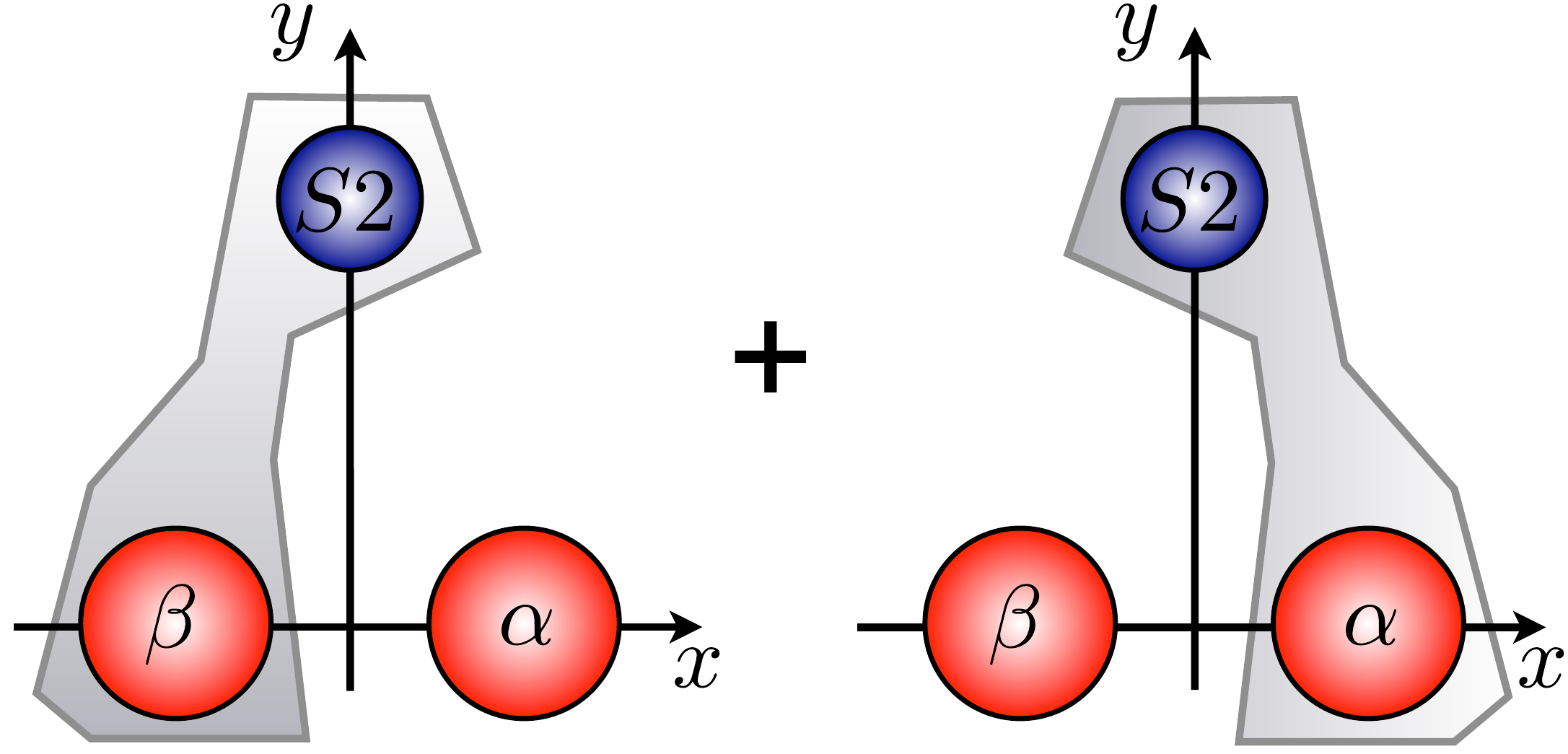}
\includegraphics[width=0.28\columnwidth]{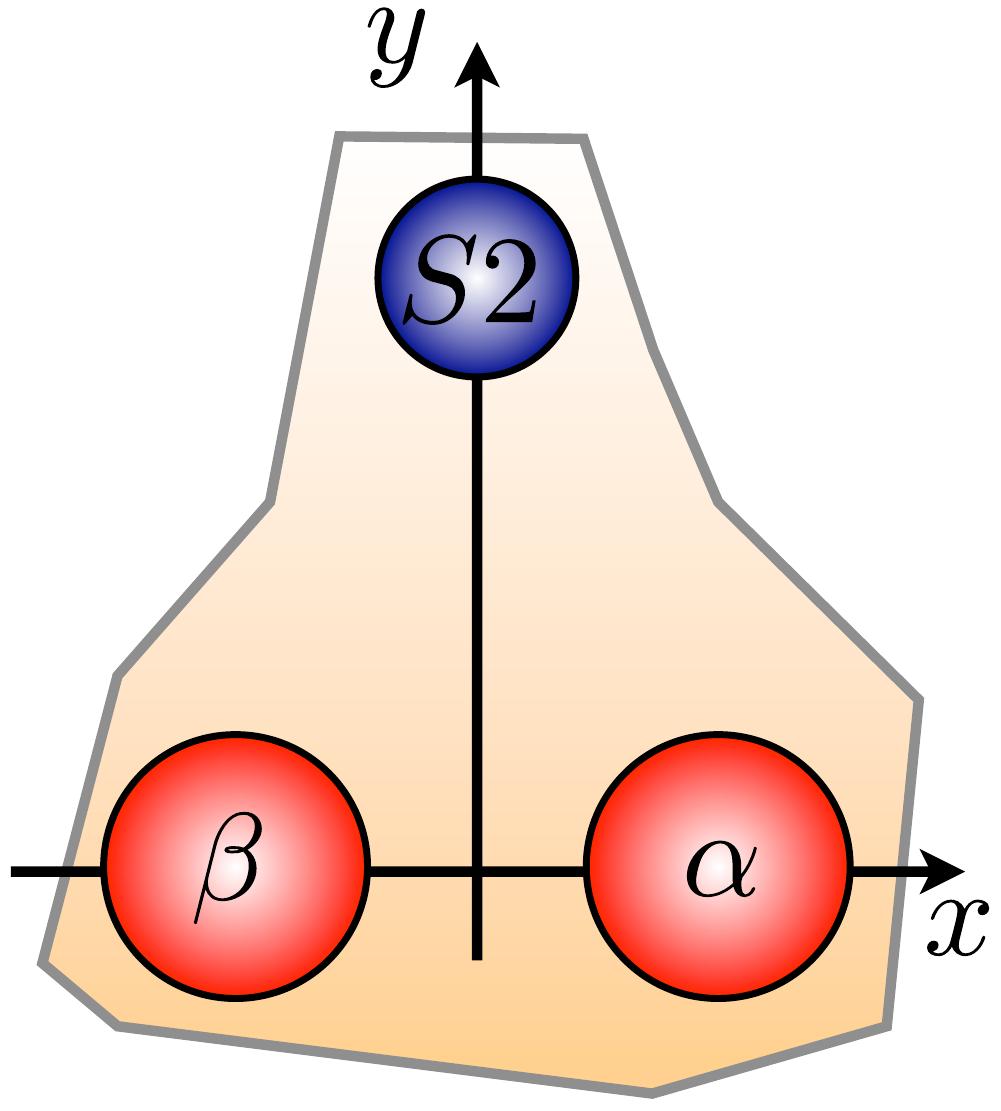}
\caption{({\bf a}) The gravitational field acting on S2 is a linear combination of gravitational fields produced by S1 being in a superposeed state. ({\bf b}) The semi-classical treatment of gravity, where the gravitational field acting on S2 is the one produced by a total mass $m_1$ with density $\frac12\left(|\alpha(x)|^2+|\beta(x)|^2\right)$.
}
\label{fig2}
\end{figure}

%%%%%%%%%%%%%%%%%%%%%%%%%%%%%%%%%%%%%%%%%%%%%%%%%%%%%%%%%%%%%%%%%%%%%%%%%%%%%%%%%%%%%5
\noindent {\it Optomechanical test}.--
\begin{figure}[t!]
\centering
\includegraphics[width=0.8\columnwidth]{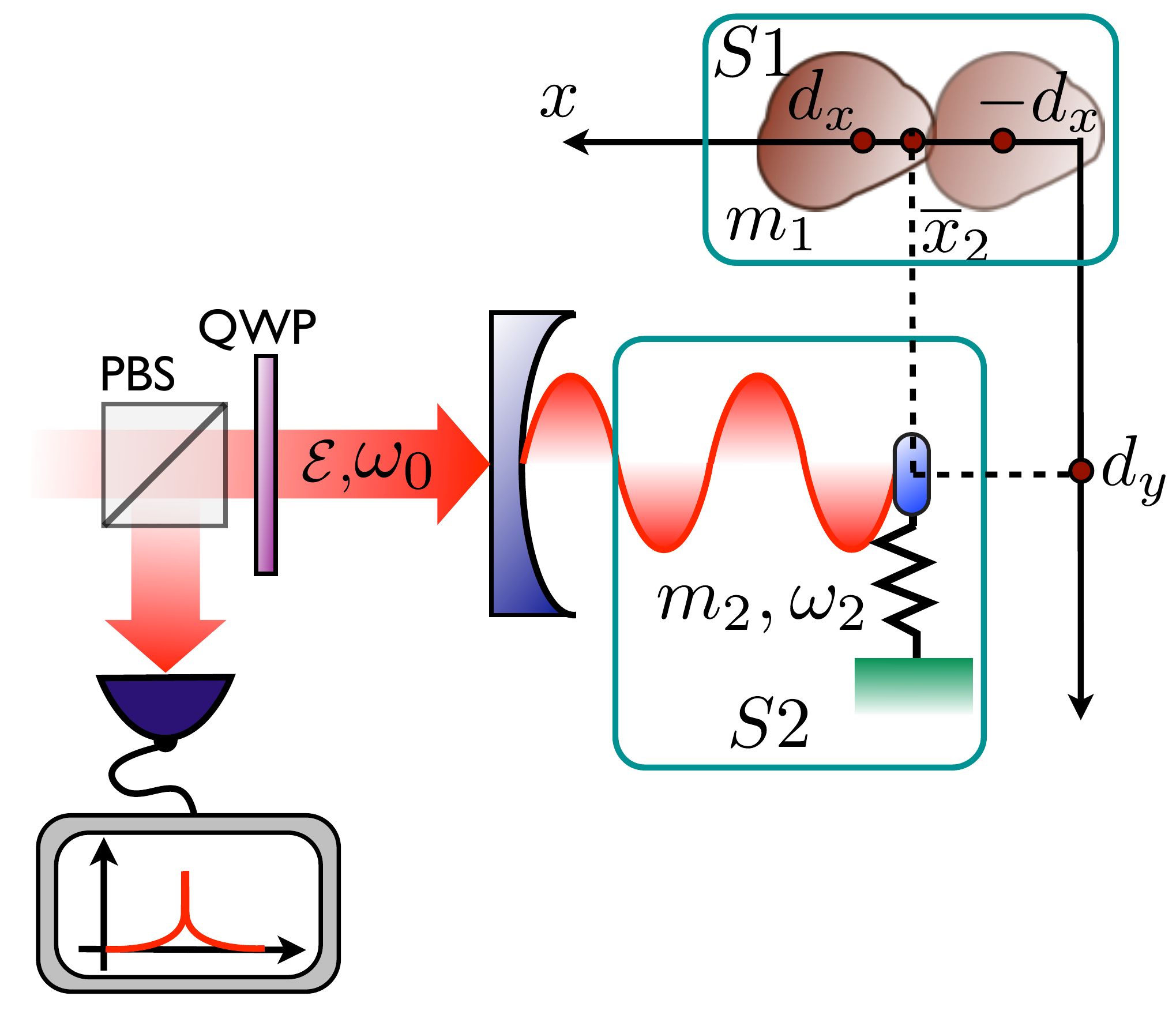}
\caption{The proposed set-up for the optomechanical falsification of quantum/classical gravity. A system S1 is prepared in a superposition of two localised states at $\pm d_x$ along the $x$ axis. An optomechanical cavity acts as transducer and a probe of (potentially quantum) gravity effects S2: the effect of the gravitational coupling between S1 and the mechanical oscillator of an optomechanical cavity induces an effect on the variance of the position fluctuations of the oscillator. The mean position of the latter along the $x$ axis is $\overline{x}_2$. The cavity is pumped by an external field (frequency $\omega_0$ and coupling rate ${\cal E}$).}
\label{fig3}
\end{figure}
We now consider the exquisite potential for motional transduction offered by optomechanics and let system S2 be the movable end-mirror of an optomechanical cavity. On the other hand, we shall not specify explicitly what is the chosen embodiment for S1, which could well be a second vibrating mechanical structure. Explicit configurations will be described elsewhere~\cite{later}. 

The transduction cavity is pumped by an external laser field and S2 is in contact with a bath of phononic modes. The axis of the optomechanical cavity, as well as the delocalization axis of S1, is assumed to lay along the $x$-axis of a reference frame. Denoting the position operator of S2 along the cavity axis by ${x}_2$, and its momentum by ${p}_2$, moving to a frame rotating at the frequency of the pumping field, we find the Hamiltonian model~\cite{Aspelmeyer}
\begin{equation}
\begin{aligned}
{H}_2&=\hbar (\omega_{c}-\omega_{0}){a}^\dagger{a}
-\hbar\,\chi\,{a}^\dagger{a}\,{x}_2\\
%\\&\nonumber
&+\frac12\,m_2\,\omega_2^2\,{x}_2^2+\frac{{p}_2^2}{2m_2}
+i\hbar {\cal E} ({a}^\dagger-{a}),
\end{aligned}
\end{equation}
where $\omega_{0}$ is the frequency of the external laser,
$\omega_{c}$ is the frequency of the cavity mode derived by the laser,
$\omega_2$ is the harmonic frequency of the mechanical oscillator,
$\chi=\omega_{c}/L$ is the optomechanical coupling constant between the cavity and the mechanical oscillator with $L$ the size of the cavity,
and ${\cal E}=\sqrt{2\kappa{\cal P}/\hbar \omega_{0}}$ with ${\cal P}$ the laser power and $\kappa$ the cavity photon decay rate.
Following conventional approach, we expand each operator as ${O}=\bar{O}+\delta{O}$ with $\bar{O}$ the steady-state mean value and $\delta{O}$ small quantum fluctuation around $\bar{O}$. Accordingly, one finds: $\bar{p}_2=0$, $\bar{x}_2=\hbar\chi|\bar{a}|^2/m_2\omega_2^2$, and $\bar{a}={\cal E}/(\kappa+i\Delta)$ with $\Delta=\omega_{c}-\omega_{0}-\chi\bar{x}_2$.
We now assume that the mean-value of the position of system S2, $\langle{\bf r}_2(t)\rangle_{\phi}$, takes a steady-state value. This implies that the coordinates of S2 in the Cartesian reference frame that we have chosen are $\langle{\bf r}_2(t)\rangle_{\phi}\approx(\bar{x}_2,d_y,0)$. Also, as $m_1\gg m_2$, within the aforementioned adiabatic approximation we have $\langle{\bf r}_1(t)\rangle_{\alpha}\approx(d_x,0,0)$ and $\langle{\bf r}_2(t)\rangle_{\beta}\approx(-d_x,0,0)$ [cf. Fig.~\ref{fig1}].
We introduce these approximations into equations of $V_\gamma$ and  $V_{\text{\tiny cl}}$ and, by taking $d_y\gg d_x,\bar{x}_2$, we find
\begin{equation}\label{V-a}
V_{\gamma} \approx -\frac{G m_1 m_2}{d_y}
\left(1+\frac{\bar{x}_2+s_\gamma d_x}{d_y^2}\,\delta x_2
-\frac{1}{d_y}\,\delta y_2\right),
%V_{\alpha} \approx& -\frac{G m_1 m_2}{d_y}
%\left(1+\frac{\bar{x}_2-d_x}{d_y^2}\,\delta x_2
%-\frac{1}{d_y}\,\delta y_2\right)\,,
%\\\label{V-b}
%V_{\beta} \approx& -\frac{G m_1 m_2}{d_y}
%\left(1+\frac{\bar{x}_2+d_x}{d_y^2}\,\delta x_2
%-\frac{1}{d_y}\,\delta y_2\right)\,,
\end{equation}
with $\gamma=\alpha,\beta$, $s_{\alpha}=-s_{\beta}=-1$, and
\begin{align}
\label{V-cl}
V_{\text{\tiny cl}} \approx 
-\frac{G m_1 m_2}{d_y} \left(1+\frac{\bar{x}_2}{d_y^2}\,\delta x_2
-\frac{1}{d_y}\,\delta y_2\right).
\end{align}
The derivative of the potentials above with respect to $\delta x_2$ contributes a term in the
quantum Langevin equation of the momentum. Accordingly, the quantum Langevin equations read
\begin{equation}
\begin{aligned}
\frac{d}{dt}\delta {x}_2(t)&=\delta{p}_2(t)/m_2,\\
\frac{d}{dt} \delta {a}(t) &= -(i\Delta+\kappa)\delta {a}(t)
+i\chi\bar{a}\,\delta {x}_2(t)+\sqrt{2\kappa}\,\delta {a}_{\text{\tiny in}}(t),\\
\frac{d}{dt} \delta {p}_2(t) &= -m_2\omega_2^2\delta {x}_2(t) + \hbar
\chi[\bar{a}\,\delta {a}^\dagger(t)+\bar{a}^*\delta {a}(t)]\\
&-\frac{\partial V_{\nu}}{\partial\, \delta x_2}
-\gamma_m\delta {p}_2(t)+{\xi}(t),
\end{aligned}
\end{equation}
where $\nu=\gamma,{\rm cl}$. Notice that $\partial V_\nu/\partial\, \delta x_2$ is not an operator-valued function. We shall denote $f=-\partial V/\partial\, \delta x_2$. Solving the above equations in the frequency domain gives us
%\begin{widetext}
\begin{equation}
\begin{aligned}
&\delta {x}_2(\omega)=-\frac{1}{D(\omega)}\{{
[\Delta^2+(\kappa-i\omega)^2][{\xi}(\omega)+2\pi f_\nu\delta(\omega)]
}\\
&{+}{i\hbar\chi\sqrt{2\kappa}\left[
\bar{a} (i\kappa+\omega{-}\Delta)\delta{a}^\dagger_{\text{\tiny in}}(\omega)
+\bar{a}^* (i\kappa+\omega{+}\Delta)\delta{a}_{\text{\tiny in}}(\omega)
\right]}\}
\end{aligned}
\end{equation}
%\end{widetext}
where $D(\omega)=m[\Delta^2+(\kappa-i\omega)^2][\omega^2-\omega_2^2+i\gamma_m\omega]
+2\hbar\chi^2|\bar{a}|^2\,\Delta$. The correlation functions of the noise operators are
\begin{equation}
\begin{aligned}
\langle\delta{a}_{\text{\tiny in}}(\omega)
\,\delta{a}^\dagger_{\text{\tiny in}}(\Omega)\rangle &=2\pi\delta(\omega+\Omega)\\
\langle{\xi}(\omega)\,{\xi}(\Omega)\rangle=
2\pi\hbar&\gamma_mm\omega\left[1+\coth(\mu\omega)\right]\delta(\omega+\Omega)
\end{aligned}
\end{equation}
with $\mu=\hbar/k_B T$. All other correlators are zero.
Therefore, one finds the spectrum of fluctuations in the position of the mechanical oscillator S2 as
%\begin{widetext}
\begin{equation}
\begin{aligned}
\label{S-omega}
{\cal S}^\nu_{x_2}&(\omega)=\frac{1}{|D(\omega)|^2}\{
2\hbar^2\chi^2\kappa|\bar{a}|^2\left(\Delta^2+\kappa^2+\omega^2\right)\\
&+\hbar\,m\,\omega\,\gamma_m\coth\left(\mu \omega\right)
[(\Delta^2+\kappa^2-\omega^2)^2+
4\kappa^2\omega^2]\}\\
&+\frac{2\pi f^2_\nu}{D(\omega)D(0)}\delta(\omega){\left(\Delta^2+\kappa^2\right)\left[\Delta^2+(\kappa-i\omega)^2\right]}.
\end{aligned}
\end{equation}
%\end{widetext}
The first two lines in this expression reproduce the standard density noise spectrum of an optomechanical system. On the other hand, the term proportional to $f^2_\nu$ in Eq.~\eqref{S-omega} is the result of the gravitational interaction. This contribution can be directly observed in the variance of the position of S2.
The variance of the fluctuations in the position of mechanical oscillator S2 is given by
\begin{align}
\langle(\delta {x}_2)^2\rangle&=\frac{1}{2\pi}
\int_{-\infty}^{+\infty}\D\omega\,{\cal S}_{x_2}(\omega).
\end{align}
Introducing Eq.~\eqref{S-omega} into the above expression yields
\begin{equation}\label{widening}
\langle(\delta {x}_2)^2\rangle=\langle(\delta {x}_2)^2\rangle_0+f^2_\nu\frac{(\Delta^2+\kappa^2)^2}{D^2(0)},%\,\left(\frac{\partial V}{\partial \delta x_2}\right)^2,
\end{equation}
where $\langle(\delta {x}_2)^2\rangle_0$ denotes the variance of the position fluctuations of the mechanical oscillator when there is no gravitational interaction. We also used $f_\nu=-\partial V_\nu/\partial\, \delta x_2$ where $V$ can be either $V_\gamma$, or $V_{\text{\tiny cl}}$ [cf. Eqs.~(\ref{V-a}) and (\ref{V-cl})]. Explicitly
\begin{align}
\label{wid-V}
\frac{\partial V_\gamma}{\partial \delta x_2}\approx-\frac{G m_1 m_2}{d_y^3}
\left(\bar{x}_2+s_\gamma d_x\right);
%
%\frac{\partial V_\alpha}{\partial \delta x_2}\approx-\frac{G m_1 m_2}{d_y^3}
%\left(\bar{x}_1-d_x\right),\, \frac{\partial V_\beta}{\partial \delta x_2}\approx-\frac{G m_1 m_2}{d_y^3}
%\left(\bar{x}_2+d_x\right).
~~
\frac{\partial V_{\text{\tiny cl}}}{\partial \delta x_2}\approx-\frac{G m_1 m_2 \bar{x}_2}{d_y^3},
\end{align}
where $\gamma=\alpha,\beta$, and $s_\alpha=-s_\beta=-1$.
As one can appreciate from Eq.~\eqref{widening}, the gravitational interaction between S1 and S2 manifests as an extra widening in the position distribution of S2. Eqs.~\eqref{widening} and (\ref{wid-V}) allow to evaluate the difference between a classical and a quantum treatment of gravity. As a figure of merit we can indeed take
\begin{align}
\Theta={Gm_1 m_2}\frac{(\Delta^2+\kappa^2)\sqrt{d_x(d_x+\bar{x}_2)}}{d^3_y|D(0)|},
\end{align}
which is the difference between the standard deviation in classical and quantum cases, and has the dimension of a length. Our goal now is to achieve the largest possible deviation. Upon inspection, one can see that $D(0)$ is minimized for $\Delta=0$ and $\omega=\omega_2$. Moreover, by assuming the (experimentally undemanding) sideband-not-resolved limit given by the condition $\kappa\gg\omega_2$, we find the {\it optimal} expression
\begin{equation}
\Theta^*=Gm_1\frac{\sqrt{d_x(d_x+\overline{x}_2)}}{d^3_y\omega_2\gamma_m}\approx  \frac{Gm_1d_x}{d^3_y\omega_2\gamma_m},
\end{equation} 
where the last expression is valid by assuming $d_x\gg\overline{x}_2$. This result shows that a high mechanical quality factor of a low-frequency oscillator would bring $\Theta^*$ to values close to observability, provided that the distance $d_y$ between the centres of mass of S1 and S2 is larger than the linear dimension of the objects along the $y$ axis. %We are now in a position to estimate the actual {\it visibility} of such difference. 
%First, as $G$ is a common scaling factor for both the classical and quantum mechanism, we consider the rescaled difference $\Theta/G$ and. 
An estimate is as follows: we take ${\cal E}=6\times10^{12}$Hz, $\gamma_m/2\pi=100$Hz, $\kappa=9\times10^7$Hz, $\omega_c/2\pi=3.7\times 10^{14}$Hz, $\omega_2/2\pi=10^7$Hz, $m_1=100$ng, %$m_2=1$ng, 
and a cavity of $1$mm length. Consistently with the formal approach above, we work under the assumptions $d_y\gg d_x,\overline{x}_2$. A suitable range of values for $d_y$ is from  $10^{-6}$m, which would be suited for micromechanical systems, to $10^{-8}$m, which would imply the use of a nanomechanical oscillator (possibly embodied by a carbon nanotube or a graphene sheet, such as in Ref.~\cite{Bachtold}). For $m_2=1$ng ($m_2=3\times 10^{-12}$g), we have $\overline{x}_2\approx7.7\times10^{-10}$m ($\overline{x}_2\approx2.6\times10^{-7}$m). For $d_y=10^{-6}$m ($d_y=10^{-8}$m ) and $d_x=5\times10^{-7}$m ($d_x=5\times10^{-9}$m), we find $\Theta^*\approx1.3\times10^{-9}G$ ($\Theta^*\approx1.3\times10^{-5}G$). The optimality of such value is assessed in Fig.~\ref{fig3}, where we show $\Theta/G$ against the frequency $\omega$ in a logarithmic plot.  As displacement sensitivities in current optomechanical devices are approaching $10^{-18}$m/$\sqrt{\rm Hz}$~\cite{Aspelmeyer,Schliesser}, the appreciation of such small deviation of the quantum treatment from the classical effects appears to be not too far from being possible, albeit technically demanding.

The small distance between S1 and S2 will make it necessary to carefully study short-range and dispersion forces such as van der Waals interactions (vdW), where strength and power law scaling of the interaction with distance dramatically depend on the actual geometry of the two interacting systems~\cite{Israel}. Competing vdW effects are notorious for investigations of rather small gravity related effects and solutions for handling vdW have been already demonstrated~\cite{ACS} .  

\begin{figure}[t!]
\centering
\includegraphics[width=1\columnwidth]{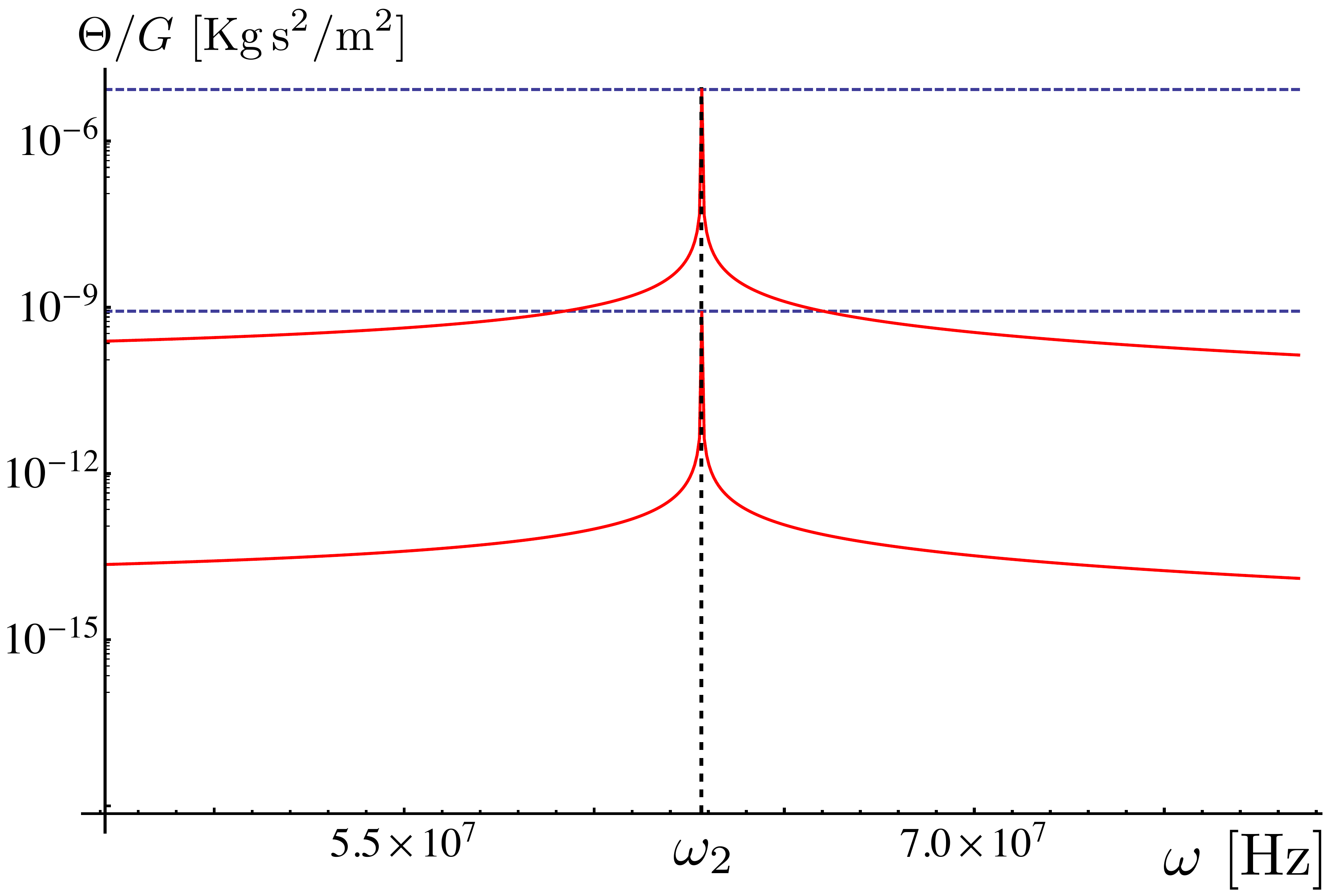}
\caption{The red lines show the plot of $\Theta$ against the frequency $\omega$ for $d_y=10^{-8}$m (lower red curve), and $d_y=10^{-6}$m with $m_2 = 1$ng (upper red curve). All other parameters as in the body of the paper. The horizontal blue lines show the values of $\Theta^*$/G corresponding to our two choices of $d_y$. We have $\max_\omega\Theta=\Theta^*$, which is achieved exactly at the mechanical frequency $\omega_2$.}
\label{fig3}
\end{figure}

%%%%%%%%%%%%%%%%%%%%%%%%%%%%%%%%%%%%%%%%%%%%%%%%%%%%%%%%%%%%%%%%%%%%%%%%%%%%%%%%%%%
\noindent{\it Conclusions}.-- We have illustrated a method to infer the {\it nature} of the gravitational interaction between two massive objects, in principle capable of discerning between a quantum and classical approach to gravity. Our approach is based on the fundamental differences occurring in light of the possibility to prepare quantum coherent states of a system, within the quantum mechanical framework, which in turn gets manifested in the possibility to achieve coherent superpositions of {\it distinguishable} gravitational fields. Such a crucial difference between classical and quantum gravity can be revealed, in principle, in an optomechanical experiment, which showcases all the necessary ingredients to falsify one of the two treatments of gravity. 

%%%%%%%%%%%%%%%%%%%%%%%%%%%%%%%%%%%%%%%%%%%%%%%%%%%%%%%%%%%%%%%%%%%%%%%%%%%%%%%%%%%
\noindent
{\it Acknowledgements.---} 
SMcM and MP thank Darren Moore and Ben Rogers for discussions during the development of this project. SMcM is grateful to the UK EPSRC for support. MB and AB acknowledge financial support from the EU project NANOQUESTFIT, %the COST Action MP1006 ``Fundamental Problems in Quantum Physics" 
and partial support from INFN. AB, MP and HU acknowledge the John Templeton Foundation (grant ID 39530 and 43467). MP and HU thank the UK EPSRC (EP/M003019/1 and EP/J014664/1). MP acknowledges the EU FP7 grant TherMiQ (Grant Agreement 618074) for financial support. HU acknowledges financial support from the Foundational Questions Institute (FQXi).

%%%%%%%%%%%%%%%%%%%%%%%%%%%%%%%%%%%%%%%%%%%%%%%%%%%%%%%%%%%%%%%%%%%%%%%%%%%%%%%%%%%

\end{document}